\documentclass[12pt]{article}
\usepackage{%
pstricks,%
pst-plot,%
epsf%
}                                  
\usepackage{cite}
\usepackage{array,dcolumn}      
\usepackage{axodraw}
\newcommand{\hs}{\hspace*{0.5cm}}
\newcommand{\be}{\begin{equation}}
\newcommand{\ee}{\end{equation}}
\newcommand{\bea}{\begin{eqnarray}}
\newcommand{\eea}{\end{eqnarray}}
\newcommand{\baa}{\begin{eqnarray*}}
\newcommand{\eaa}{\end{eqnarray*}}

\begin{document}
\begin{center}
{\large \bf Anomalous magnetic moment of muon in 3 - 3 - 1 models}
\vspace*{1cm}

{\bf Nguyen Anh  Ky  $^{a,}$}\footnote{On leave from
 Institute of Physics, NCST, P. O. Box 429, Bo Ho, Hanoi 10000,
Vietnam}
, ~{\bf  Hoang Ngoc Long $^{b,1}$ },
and   {\bf  Dang Van Soa $^{c,}$}\footnote{On leave  from
 Department of Physics, Hanoi University of Mining and Geology,
Hanoi, Vietnam}\\
\vspace*{0.3cm}

\medskip

$^a$ {\it Department of Physics, Chuo University,
Tokyo 112 - 8551,  Japan}\\
$^b$ {\it Fakult\"at f\"ur Physik, Universit\"at Bielefeld,
D-33615 Bielefeld, Germany} \\
$^c$ {\it The Abdus Salam International
Centre for Theoretical Physics,
Trieste, Italy}\\
\vspace*{1cm}
{\bf  Abstract}\\
\end{center}

\hs A contribution from new gauge bosons in the
$\mbox{SU}(3)_C\otimes \mbox{SU}(3)_L \otimes 
\mbox{U}(1)_N$ (3 - 3 - 1) models to the anomalous 
magnetic moments of the muon is calculated and 
numerically  estimated. In the minimal 3 - 3 - 1 model,
a lower bound on the bilepton mass at a value of 167 GeV is derived.
For an expected precision($\sim 4\times 10^{-10}$) of the BNL 
measurements the possible lower bounds on masses of the bileptons
in the minimal version and in the version with right-handed
neutrinos are around 940 GeV and 250 GeV, respectively.\\[5mm]

PACS number(s): 13.40.Em, 14.60.Ef, 14.70.Pw

\section{Introduction}
\hs The SuperKamiokande results~\cite{suk} confirming
non-zero neutrino mass call for the standard model (SM)
extension.
Among the known extensions, the models based on the
$\mbox{SU}(3)_C\otimes \mbox{SU}(3)_L \otimes \mbox{U}(1)_N$ 
gauge group~\cite{ppf,fhpp} have the following intriguing features:
firstly, the models are anomaly free only if the number
of families $N$ is a multiple of three. Further,
from the condition of QCD asymptotic freedom,
which means $N < 5$, it follows that $N$
is equal to 3. The second characteristic
is that  the Peccei--Quinn~\cite{pq} symmetry, a solution
of the strong CP problem  naturally occurs in these
models~\cite{pal}.
The third interesting feature
is that one of the quark families  is treated differently
from the other two~\cite{jl,fr95}. This could lead to a natural
explanation of the unbalancing heavy top quarks in  the fermion
mass hierarchy~\cite{fr95}. Recent analyses have indicated that
signals of new particles in this model, bileptons~\cite{can} and exotic
quarks~\cite{jm} may be observed at the Tevatron and the Large
Hadron Collider (LHC).\par
\bigskip

\hs There are two main versions of the 3 - 3 - 1 models: the minimal
 model in which all lepton components $(\nu, l, (l)^c)_L$ of each
family belong to one and same lepton triplet and a variant, in
which right--handed neutrinos (r. h. neutrinos) are included,
i.e. $(\nu, l, \nu^c)_L$ 
(hereafter we call it a model with right-handed neutrino
~\cite{rhnm,mpp}).
New gauge bosons in the minimal model are  bileptons
 ($Y^\pm, X^{\pm\pm}$) carrying lepton number $L = \pm 2$ and $Z'$.
In the second model, the bileptons with lepton number $L = \pm 2$
are singly--charged $Y^\pm$ and {\it neutral} gauge bosons
$X^0, X^{*0}$ , and both are responsible for lepton--number
violating interactions. \\

\hs With the present group extension there are 
five new gauge bosons and all
these particles are heavy. Getting mass limits for these particles
is one of the central tasks of further studies. The anomalous
magnetic moments
of the muon (AMMM) $a_\mu \equiv (g_\mu -2)/2$ is one of the
most popular values in pursueing this aim. Despite not competitive
with the anomalous magnetic moment of the electron (AMME) in
precision, the AMMM is much more sensitive to loop effects as well
as``New Physics" due to contributions $\sim m_\mu^2$, i.e.
$\sim (200)^2$ enhancement in the AMMM relative to the AMME.
 Therefore the AMMM is a subject of both theoretical
and experimental investigations~\cite{mar}.
The $(g_\mu -2)/2$ was used to get constraints on mass of
the bilepton in the minimal version~\cite{fnsasaki}. However in the
cited paper a contribution from new neutral gauge
boson $Z'$ was not included.\\

\hs The aim of this work is to calculate  the $(g_\mu -2)/2$ in both
3 - 3 - 1 versions. As a consequence,  constraints on the new
gauge boson masses are discussed.\\

\hs Our paper is organized as follows: In Sec. 2 after a brief
introduction into the minimal version, we present contributions
to the $(g_\mu -2)/2$ from both bileptons and $Z'$. Constraints
on their masses are also  derived. Sec. 3 is devoted
to the version with r.h. neutrinos. Finally, our
conclusions are summarized in the last section.

\section{The $(g_\mu -2)/2$ in the minimal version }
\hs Let us firstly  recapitulate the basic
elements of the model (for more details see~\cite{dng}).
Three lepton components of each family  are in one triplet:
\be
f^{a}_L = \left(
               \nu^a,\  l^a,\ (l^c)^a
                 \right)_L^T \sim (1, 3, 0),
\label{l}
\ee
where $ a = 1, 2, 3$ is the family index. The charged bileptons
with lepton number $L = \pm 2$ are identified as follows:
$\sqrt{2}\ Y^-_\mu = W^4_\mu- iW^5_\mu ,
\sqrt{2}$\ $ X^{--}_\mu =
W^6_\mu- iW^7_\mu $, and their
couplings  to leptons are given by~\cite{framcal}
\be
{\cal L}^{CC}_l = - \frac{g}{2\sqrt{2}}\left[
\bar{\nu}\gamma^\mu (1- \gamma_5) C\bar{l}^{T}Y^-_\mu
 -  \bar{l}\gamma^\mu \gamma_5 C \bar{l}^T
X^{--}_\mu + \mbox{h.c.}\right].
\ee
It is to be noted that the vector currents coupled to $X^{--}$,
$X^{++}$ vanish due to Fermi statistics. To get physical neutral
gauge bosons one has to  diagonalize  their mass mixing
matrix. That can be done in two steps:
At the first, the photon field $A_\mu$ and $Z,Z'$ are given
by~\cite{dng}
\bea
A_\mu  &=& s_W  W_{\mu}^3 + c_W\left(\sqrt{3}\  t_W  W^8_{\mu}
+\sqrt{1- 3\ t^2_W}\ B_{\mu}\right),\nonumber\\
Z_\mu  &=& c_W  W_{\mu}^3 - s_W\left(\sqrt{3}\  t_W  W^8_{\mu}
+\sqrt{1- 3\ t^2_W}\ B_{\mu}\right),\nonumber\\
Z'_\mu & = & \sqrt{3}\  t_W \  B_{\mu}
- \sqrt{1- 3\ t^2_W}\ W^8_{\mu},
\label{apstat}
\eea
where, as  usual, the notation  $s_W \equiv
\sin\theta_W$ is used.
In the second step, we get the physical neutral gauge bosons
$Z^1$ and $Z^2$ which are mixtures of  $Z$ and $ Z'$:
\bea
Z^1  &=&Z\cos\phi - Z'\sin\phi,\nonumber\\
Z^2  &=&Z\sin\phi + Z'\cos\phi.
\eea
The mixing angle $\phi$ is constrained to be very small, therefore
the $Z$ and the $Z'$ can be safely  considered as the physical
particles.

\hs The gauge interactions for  $Z'$ can be written in the form
\be
{\cal L}^{NC} = \frac{g}{c_W}\left\{\bar{f}\gamma^{\mu}
[g'_{V}(f) + g'_{A}(f)\gamma_5]f Z'_{\mu}\right\}.
\label{ncm}
\ee
The alternative left--right form with coupling
coefficients~\cite{dng}
\be
g'_{L,R}(f) = - \frac{\sqrt{1 - 4 s_W^2}}{2\sqrt{3}}Y(f_{L,R})
 + \frac{1-s_W^2}{\sqrt{3 (1 - 4s_W^2)}} N(f_{L,R}),
\label{vazm}
\ee
has simple relations
$g'_{V}(f) = [g'_{R}(f)+ g'_{L}(f)]/2$,
$g'_{A}(f) = [g'_{R}(f)- g'_{L}(f)]/2$.

\hs Now we calculate contributions from the bileptons and the
$Z'$ to the AMMM. It is known that heavy Higgs boson contribution
to the AMMM is negligible~\cite{jw}, therefore
the relevant diagrams are depicted in Fig.1.
\vspace*{0.5cm}

The first three diagrams come from the bileptons and their
contributions are found to be
\be
\delta a_\mu^B = \frac{g^2 m_\mu^2}{24 \pi^2} \left(
\frac{16}{M_X^2} + \frac{5}{4 M_Y^2} \right),
\label{dgbm}
\ee
where $M_X$, $M_Y$, $m_\mu$ stand for  masses of the doubly-,  
singly-charged bileptons and of the muon, respectively.
In the limit $m_\mu << M_{Z'}$ where $M_{Z'}$ is the $Z'$ mass,
 the $Z'$ contribution has the form~\cite{moore}
\be
\delta a_\mu^{Z'} = \frac{m_\mu^2}{12 \pi^2 M_{Z'}^2}
 \left(g^{'2}_{V} - 5 g^{'2}_{A}\right).
\label{dgz}
\ee
Applying  Eq.(\ref{vazm}) we get coupling of the muon to  the $Z'$
\be
g'_{V}(\mu) = \frac{g}{c_W} \frac{3\sqrt{1 - 4 s_W^2}}{2\sqrt{3}},\
g'_{A}(\mu) = \frac{g}{c_W} \frac{\sqrt{1 - 4 s_W^2}}{2\sqrt{3}}.
\label{hsm}
\ee
Substituting (\ref{hsm}) into (\ref{dgz}) we obtain the $Z'$
contribution
\begin{equation}
\delta a_\mu^{Z'} = \frac{g^2}{3 c_W^2}
 \frac{m_\mu^2}{12 \pi^2 M_{Z'}^2} ( 1-4 s_W^2 ).
\label{dgzm}
\ee
Therefore the total contribution from  new gauge bosons in
the minimal version to the AMMM becomes
\be
\delta a_\mu^{tm} = \frac{G_F m_W^2 m_\mu^2}{3\sqrt{2}\pi^2}
\left[ \frac{16}{ M_X^2} + \frac{5}{4 M_Y^2} +
 \frac{2 (1-4 s_W^2)}{3 c_W^2
M_{Z'}^2} \right],
\label{dgzmt}
\ee
where $G_F/\sqrt{2} = g^2/(8 m_W^2)$ is used.

\hs Note that the $Z'$ gives a positive contribution to the AMMM,
while the $Z$ gives a negative one as it is well--known in the SM.
From Eq. (\ref{dgzmt}) it follows that the bilepton contributions
are dominant.

\hs By the spontanous symmetry breaking (SSB) it
follows that~\cite{lng}
$|M_X^2 - M_Y^2| \leq {\cal O}(m_W^2)$
(more precisely $|M_X^2 - M_Y^2| \leq 3 m_W^2$) .
Therefore it is acceptable
to put $M_X \sim M_Y$ as it was done in~\cite{fnsasaki}.
In this approximation,  Eq. (\ref{dgbm}) agrees with the original
result in~\cite{fnsasaki}, and Eq. (\ref{dgzmt}) becomes
\be
\delta a_\mu^{tm} = \frac{G_F m_W^2 m_\mu^2}{\sqrt{2}\pi^2}
\left[ \frac{23}{4 M_Y^2} +
 \frac{2 (1-4 s_W^2)}{9 c_W^2
M_{Z'}^2} \right].
\label{dgzmtt}
\ee

\hs  A lower limit  $M_Y \sim 230$ GeV at  95\% CL  can
be extracted by  the ``wrong" muon decay $\mu
\rightarrow e^- \nu_e \bar{\nu}_\mu $ .
Combining with the SSB, it
follows~\cite{dng} $M_{Z'} \geq 1.3$ TeV.
With the quoted numbers ($ M_X = 180, M_Y = 230, M_{Z'} = 1300$ GeV),
the contributions to $\delta a_\mu^{tm}$ from the bileptons
and the $Z'$ are $1.04 \times 10^{-8}$ and
$7.76 \times 10^{-13}$, respectively. The bilepton contribution
is in a range of ``New Physics" one~\cite{cza}
 $\sim {\cal O}(10^{-8})$.

\hs Putting a bound on ``New Physics" contribution to the
AMMM~\cite{ki}
\be
\delta a_\mu^{New\  Physics} = (7 \pm 8.6) \times 10^{-9},
\label{np}
\ee
into the l.h.s of (\ref{dgzmtt}) we can obtain
a bound on $M_{Y}$. In Fig. 2 we plot $\delta a_\mu^{tm}$
as a function of $M_Y$.
For certainty we used $M_{Z'} = 1.3$ TeV quoted above.
The horizontal lines are the upper and the lower  limit from
$\delta a_\mu^{New\  Physics}$.
\vspace*{0.5cm}

 From the figure we get a lower mass limit on $M_{Y}$ to
be 167 GeV.
We recall that this limit is in a range of  those
obtained from LEP data analysis ($M_Y \ge 120$ GeV)~\cite{fram}.

\hs In the near future, the E-821 Collaboration at Brookhaven would
reduce the experimental error on the AMMM to a few
$\times 10^{-10}$. 
\vspace*{1cm} 

In Fig. 3 we see that $\delta a_\mu^{tm}$
cuts horizontal line I ($\sim 4 \times 10^{-10}$) and line II
($\sim 1 \times 10^{-10}$) at $M_Y \approx 935$ GeV and 
 $M_Y \approx 1870$ GeV, respectively. These lower bounds 
are much higher than those from the muon experiments.\\[0.3cm]

\section{The $(g_\mu -2)/2$ in the model with r.h. neutrinos}
\hs In this version the third member of the lepton
triplet is a r. h. neutrino instead of the antilepton
$l^c_L$
\be
f^{a}_L = \left(  \nu^a, l^a,  (\nu^c)^a
\right)_L^T \sim (1, 3, -1/3), l^a_R\sim (1, 1, -1).
\label{lr}
\ee
The complex new gauge bosons
$\sqrt{2}\ Y^-_\mu = W^6_\mu- iW^7_\mu ,\sqrt{2}\ X^0_\mu
=W^4_\mu- iW^5_\mu $ are responsible for lepton--number
violating interactions. Instead of the  doubly--charged
bileptons $X^{\pm\pm}$, here  we have  neutral
ones $X^0$, $X^{0*}$. The SSB gives
the bilepton  mass splitting~\cite{li}
\[ |M_Y^2 - M_X^2| \leq m_W^2.\]

\hs As before one diagonalizes
the mass mixing  matrix  of the neutral gauge
bosons by  two steps, and the last one
is the same for both versions. At the first step we have
\bea
A_\mu  &=& s_W  W_{\mu}^3 + c_W\left(-
\frac{t_W}{\sqrt{3}}\ W^8_{\mu}
+\sqrt{1-\frac{t^2_W}{3}}\  B_{\mu}\right),\nonumber\\
Z_\mu  &=& c_W  W^3_{\mu} - s_W\left(
-\frac{t_W}{\sqrt{3}}\ W^8_{\mu} +
\sqrt{1-\frac{t_W^2}{3}}\  B_{\mu}\right), \nonumber \\
Z'_\mu &=& \sqrt{1-\frac{t_W^2}{3}}\  W^8_{\mu}
+\frac{t_W}{\sqrt{3}}
\ B_{\mu}.
\label{apstatr}
\eea
Due to smallness of mixing angle $\phi$ we can consider
the $Z$ and  the  $Z'$ as the physical particles. The couplings of
fermions with
$Z'$ boson are given as follows~\cite{lv}:
\be
g'_{L,R}(f) =  c_W^2\left[\frac{3N(f_{L,R})}{(3-4s_W^2)^{1/2}}
-\frac{(3-4s_W^2)^{1/2}}{2c^2_W}Y(f_{L,R})\right].
\label{vazr}
\ee

From (\ref{vazr}), the couplings of $Z'$ to muon are found to be
\be
g'_{V}(\mu) = \frac{g}{4 c_W} \frac{(1 -
4 s_W^2)}{\sqrt{3-4s_W^2}},\
g'_{A}(\mu) = -\frac{g}{4 c_W \sqrt{3-4s_W^2}}.
\label{hsr}
\ee

Due to its neutrality, the  bilepton  $X^0$ does not
give a contribution and in this case, the relevant diagrams are
only two last (c) and (d). The contribution from the
singly--charged bilepton and the $Z'$ in Fig. 1(c) and 1(d) is
\be
\delta a_\mu^{tr} = \frac{G_F m_W^2 m_\mu^2}{12\sqrt{2}\pi^2}
\left\{ \frac{5}{M_Y^2} -  \frac{[5-(1-4 s_W^2)^2]}{2 c_W^2
(3 - 4 s_W^2)M_{Z'}^2} \right\}.
\label{dgzrt}
\ee
 In the considered version the $Z'$ gives a negative contribution.
However, the total value in r.h.s of Eq. (\ref{dgzrt}) is
positive (an opposite sign happens when $M_{Z'} \leq 0.3\  M_Y$
which is excluded by the SSB).

\hs Putting the $Z'$ lower mass bound  to be 1000 GeV~\cite{lv}
 followed from $\Delta m_K $ and  $M_Y = 230$ GeV we get
the bilepton and the $Z'$ contributions to
$\delta a_\mu^{tr}$, respectively: $4.75 \times 10^{-10}$ and
$- 7.87 \times 10^{-12}$. This implies that the contribution
of the new gauge bosons in the considered version is in two order
 smaller than an allowed difference between theoretical calculation
in the SM and present experimental precision.
\vspace*{1cm}

\hs However, putting two previous values for $\delta a_\mu^{tr}$
we get lower bounds on the bilepton masses to be about 250 GeV (I)
and 500 GeV (II) (see Fig. 4).\\[0.3cm]

\section{Conclusion}

\hs In conclusion, we have calculated in detail the
second--order contribution from new gauge bosons in the 3 - 3 - 1
models to the AMMM.
 In the minimal version, contribution from the $Z'$
is positive but suppressed due to a factor $(1- 4 s_W^2)$ while
in the version with r.h.neutrinos the contribution $Z'$ is negative.
In both cases the bilepton contribution is bigger by an absolute
value and the total contribution from the new gauge bosons is
positive. Comparing with experimental bounds on the AMMM
we get the lower bounds on the bilepton mass:
$M_{Y} > 167$ GeV in the minimal version.
For  the version with r.h.neutrinos the contribution
of the new gauge bosons is in two order smaller than those
in the minimal one and it does not allow to get a constraint
at the present status on the AMMM.

\hs Although analysis on the AMMM could not give a better limit
on the bilepton  mass than those from other studies~\cite{bi},
the study on the
contribution of new gauge bosons to the AMMM is in its own
right very important. However our limit can be made more restrictive
by including further experiments.

\hs With the expected experimental error on the AMMM 
at BNL to be a few $\times 10^{-10}$ the lower bounds on masses of 
the bileptons in the minimal and in the version with r.h. neutrinos
are around 1 TeV and 400 GeV, respectively.

\bigskip

{\bf Acknowledgments}

\hs The work of H. N. L. was supported by the DAAD fellowship.
He thanks Prof. D. Schildknecht and Theory Group, Bielefeld
University for hospitality and support. N.A.K. would like to 
thank T. Inami and Department of Physics, Chuo University, Tokyo, 
Japan, for warm hospitality and the Nishina memorial foundation 
for financial support.\par
\hs This work was supported in part by the 
Research Programme on Natural Science 
of Hanoi National University 
under grant number QT 98.04 and KT -04.1.2.

\newpage
\markright{FIGURES}
\begin{center}
\begin{picture}(260,50)(-5,0)
\Photon(1,20)(1,45){2}{3}
\ArrowLine(1,20)(-20,-10)
\ArrowLine(22,-10)(1,20)
\Photon(-20,-10)(22,-10){2}{5}
\ArrowLine(22,-10)(43,-40)
\ArrowLine(-41,-40)(-20,-10)
\Text(1,-45)[]{(a)}
\Text(10,30)[]{$\gamma$}
\Text(-18,10)[]{$\mu^-$}
\Text(-38,-20)[]{$\mu^-$}
\Text(26,10)[]{$\mu^-$}
\Text(46,-20)[]{$\mu^-$}
\Text(0,-20)[]{$X^{--}$}

\Photon(185,20)(185,45){2}{3}
\Photon(185,20)(168,-10){2}{4}
\Photon(185,20)(202,-10){2}{4}
\ArrowLine(202,-10)(168,-10)
\ArrowLine(151,-40)(168,-10)
\ArrowLine(202,-10)(219,-40)
\Text(185,-45)[]{(b)}
\Text(195,30)[]{$\gamma$}
\Text(165,10)[]{$X^{--}$}
\Text(151,-20)[]{$\mu^-$}
\Text(212,10)[]{$X^{- -}$}
\Text(227,-20)[]{$\mu^-$}
\Text(185,-18)[]{$\mu^{-}$}
\Photon(1,-80)(1,-105){2}{3}
\Photon(1,-105)(-19,-140){2}{4}
\Photon(1,-105)(22,-140){2}{4}
\ArrowLine(22,-140)(-19,-140)
\ArrowLine(22,-140)(43,-170)
\ArrowLine(-40,-170)(-19,-140)
\Text(1,-175)[]{(c)}
\Text(10,-100)[]{$\gamma$}
\Text(-19,-120)[]{$Y^-$}
\Text(-34,-150)[]{$\mu^-$}
\Text(27,-120)[]{$Y^-$}
\Text(45,-150)[]{$\mu^-$}
\Text(1,-149)[]{$\nu_\mu$}
\Photon(185,-80)(185,-105){2}{3}
\ArrowLine(164,-140)(185,-105)
\ArrowLine(185,-105)(206,-140)
\Photon(164,-140)(206,-140){2}{5}
\ArrowLine(145,-170)(164,-140)
\ArrowLine(206,-140)(224,-170)
\Text(185,-175)[]{(d)}
\Text(195,-100)[]{$\gamma$}
\Text(165,-120)[]{$\mu^-$}
\Text(150,-150)[]{$\mu^-$}
\Text(210,-120)[]{$\mu^-$}
\Text(227,-150)[]{$\mu^-$}
\Text(185,-150)[]{$Z'$}
\Text(87,-200)[]{ Figure  1}
\end{picture}
\end{center}
\newpage
\input{fig2.tex}
\input{fig3.tex}
\newpage 
\input{fig4.tex}
\begin{center}
\vspace*{2cm}
{\bf  FIGURE CAPTIONS}
\end{center}
\begin{itemize}
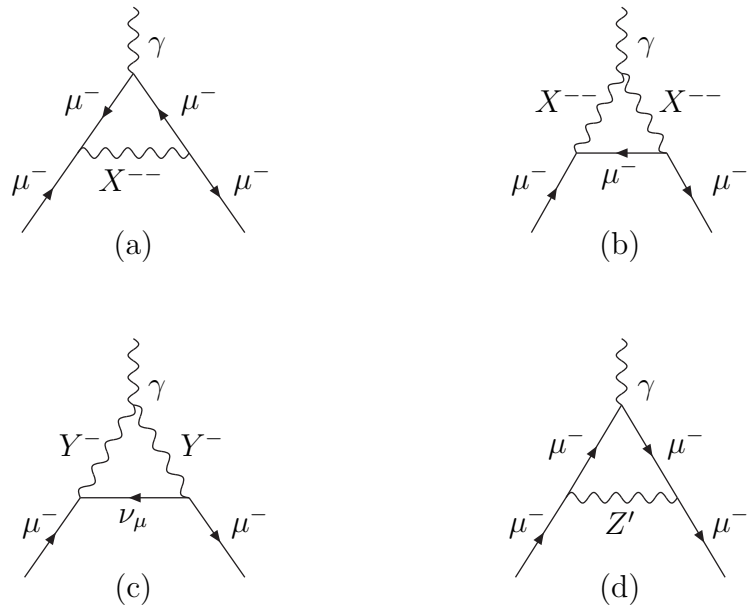
\item Fig. 1: Diagrams for the $(g_\mu -2)/2$\\
-- (a), (b), (c), (d) in the minimal version\\
-- (c), (d) in the version with r.h. neutrinos
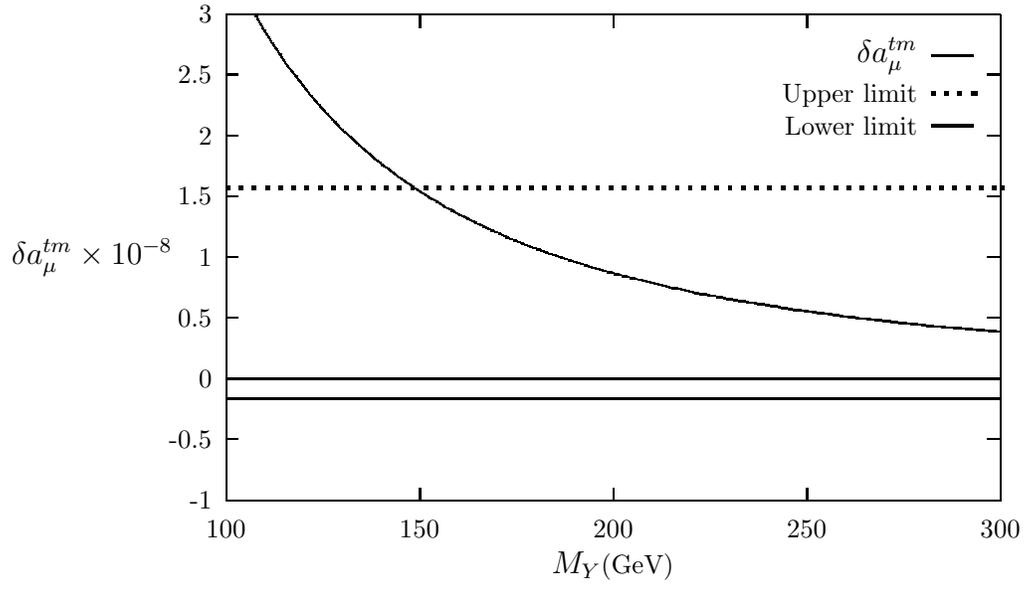
\item  Fig. 2:   $\delta a_\mu^{tm}$ as a function of $M_{Y}$,
where $M_{Z'}= 1300 $ GeV is used.
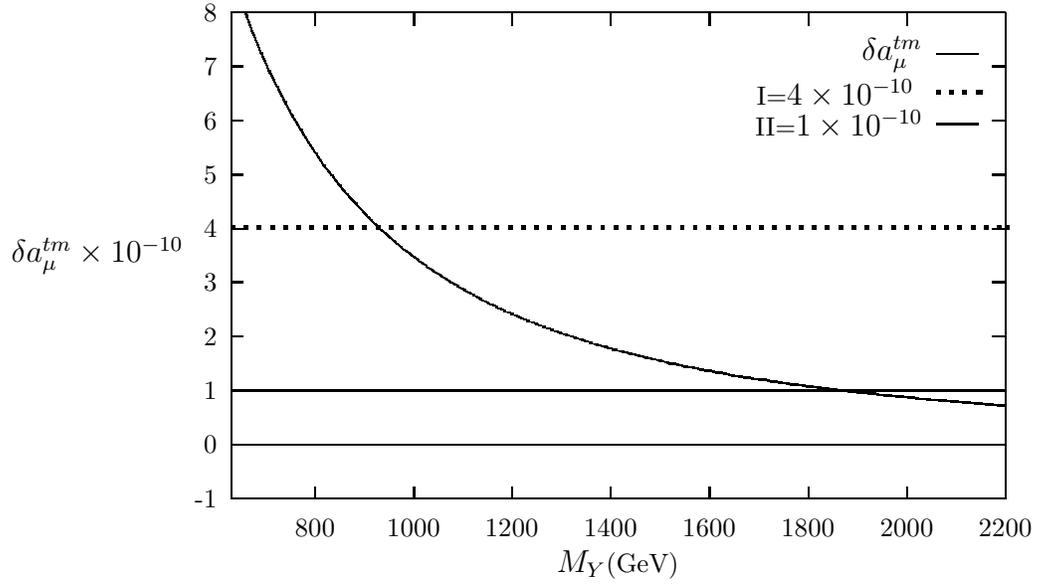
\item  Fig. 3:   $\delta a_\mu^{tm}$ as a function of $M_{Y}$,
where $M_{Z'}= 1300 $ GeV is used. Here I and II are 
two expected BNL  experimental values.
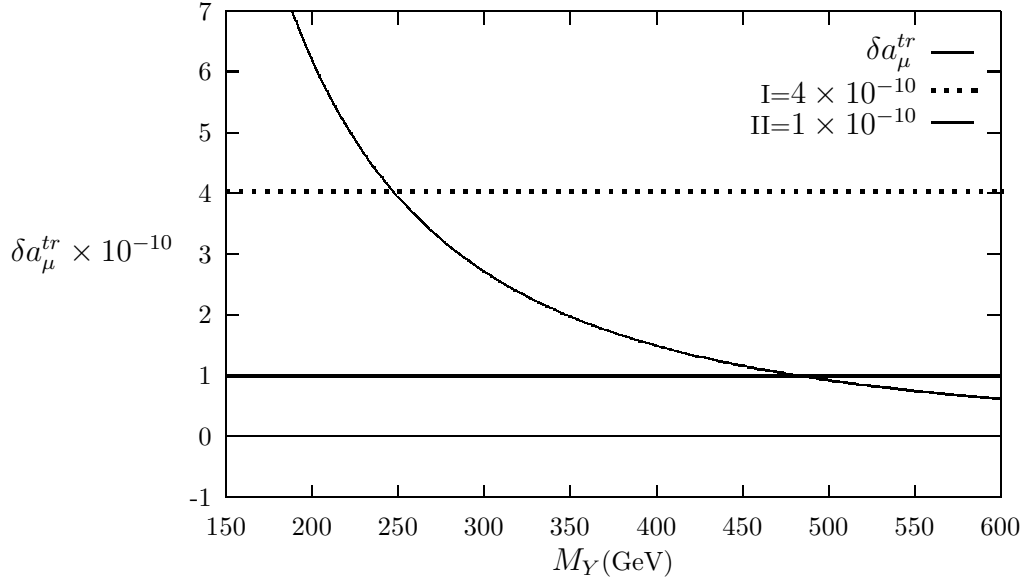
\item  Fig. 4:   $\delta a_\mu^{tr}$ as a function of $M_{Y}$,
where $M_{Z'}= 1000 $ GeV is used. Here I and II are 
two expected BNL  experimental values.
\end{itemize}

\begin{thebibliography}{99}
\bibitem{suk}Y. Fukuda {\it et al.}, Phys. Lett.  B 433 (1998) 9;
Phys. Rev. Lett.  81 (1998) 1562; T. Kajita,
{\it in} Proceedings of the XVIIIth International
Conference on Neutrino Physics and Astrophysics, Takayama,
 Japan (June 1998).
\bibitem{ppf} F. Pisano and V. Pleitez, Phys. Rev. D 46
(1992) 410;
P. H. Frampton, Phys. Rev. Lett.  69 (1992) 2889.
\bibitem{fhpp} R. Foot, O.F. Hernandez, F. Pisano, and
V. Pleitez, Phys. Rev. D  47 (1993) 4158.
\bibitem{pq}R. D. Peccei and H. R.  Quinn, Phys. Rev. Lett.
 38 (1977) 1440; Phys. Rev. D 16  (1977) 1791.
\bibitem{pal}P. B. Pal, Phys. Rev. D 52 (1995) 1659.
\bibitem{jl}J. T. Liu, Phys. Rev. D  50 (1994) 542.
\bibitem{fr95}P. H. Frampton, {\it in} Proceedings of
Workshop on Particle Theory and Phenomenology, International
Institute of Theoretical and Applied Physics, Ames, Iowa (1995),
[hep-ph/9507351].
\bibitem{can}B. Dion, T. Gregoire, D. London, L. Marleau, and
H. Nadeau, [hep-ph/9810534], Phys. Rev. D 59 (1999) 075006.
\bibitem{jm}P. Das, P. Jain, and D. W. Mckay, [hep-ph/9808256],
Phys. Rev. D  59 (1999) 055011; Y. A. Coutinho, P. P.
Queroz, and M. D. Tonasse, [hep-ph/9907553], Phys. Rev. D  60
 (1999) 115001.
\bibitem{rhnm}R. Foot, H. N. Long, and Tuan A. Tran,
Phys. Rev. D  50 (1994) R34; H. N. Long, {\it ibid} D 53 (1996) 437;
{\it ibid}. D  54 (1996) 4691.
\bibitem{mpp}J. C. Montero, F. Pisano, and V. Pleitez, Phys. Rev.
D  47 (1993) 2918.
\bibitem{mar}See for example {\it Quantum Electrodynamics},
edited by T. Kinoshita (World Scientific, Singapore) 1990.
\bibitem{fnsasaki}H.~Fujii, S.~Nakamura, and K.~Sasaki,
Phys. Lett.  B 299 (1993) 342.
\bibitem{dng}D. Ng, Phys. Rev.  D  49 (1994) 4805.
\bibitem{framcal}P. H. Frampton and D. Ng, Phys. Rev. D  45
 (1992) 4240.
\bibitem{jw}See for example R. Jackiw and S. Weinberg, Phys. Rev.
D  5 (1972) 2473; G. Altarelli, N. Cabibbo, and L. Maiani,
Phys. Lett.  B 40 (1972) 415; W. A. Bardeen, R. Gastmans,
and B. E. Lautrup, Nucl. Phys.  B 46 (1972) 319.
\bibitem{moore}For details see, S. R. Moore, K. Whisnant,
and B.-L. Young, Phys. Rev. D  31 (1985) 105;
J. P. Levelle, Nucl. Phys.  B 137 (1978) 69.
\bibitem{lng}J. T. Liu and D. Ng, Z. Phys.  C 62 (1994) 693.
\bibitem{cza}A. Czarnecki and W. J. Marciano,
BNL-HET-98-43.
Talk given at 5th International Workshop on
Tau Lepton Physics (TAU 98), Santander, Spain, 14-17 Sep 1998.
In *Santander 1998, Tau'98* 245-252, [ hep-ph/9810512].
\bibitem{ki}T. Kinoshita, Talk given at the International
Symposium on  {\it Lepton moments},
June 8 - 12, 1999, Heidelberg, Germany.
[http://www.physi.uni-heidelberg.de/$^{\sim}$muon/lep/lep.html].
\bibitem{fram}P. Frampton, Int. J. Mod. Phys.  A 13 (1998) 2345.
\bibitem{li}H. N. Long and T. Inami, [hep-ph/9902475],
Phys. Rev. D  61 (2000) 075002.
\bibitem{lv}H. N. Long and V. T. Van, [hep-ph/9909302]
J. Phys. G  25 (1999) 2319.
\bibitem{bi}For the recent discussion on bilepton mass see,
L. Willmann {\it et al}, Phys. Rev. Lett.  82 (1999) 49;
M. B. Tully and G. C. Joshi, Int. J. Mod. Phys.  A 13  (1998) 5593;
UM-P-99/16,[hep-ph/9905552] Phys. Lett.  B 466 (1999) 333;
 P. Das and P. Jain, [hep-ph/9903432];
V. Pleitez, [hep-ph/9905406] Phys. Rev. D  61 (2000) 055903.
\end{thebibliography}
\end{document}